\documentclass[aps,twocolumn,nofootinbib,preprintnumbers,prl,10pt]{revtex4-2}
\usepackage[utf8]{inputenc}

\usepackage[svgnames]{xcolor}
\usepackage[most]{tcolorbox}
\usepackage{caption}
\tcbset{colback=PaleGreen!0!white, colframe=NavyBlue, 
	highlight math style= {enhanced, 
		colframe=red,colback=red!10!white,boxsep=0pt}
}
\usepackage{relsize}

\usepackage{amsmath}
\usepackage[bottom]{footmisc}
\usepackage{braket}
\usepackage{graphicx}
\usepackage{mwe}
\usepackage[section]{placeins}
\usepackage{framed}
\usepackage{csquotes}
\usepackage{tikz}
\usetikzlibrary{decorations.markings}
\usetikzlibrary{decorations.pathmorphing}
\definecolor{shadecolor}{rgb}{0.90,0.90,0.90}
\usepackage{hyperref}
\usepackage{bbold}
\usepackage{subcaption}
\usepackage{pifont}
\usepackage{setspace}
\usepackage{amsmath, amssymb, amsthm, float, graphicx,amsfonts}
\usepackage{amsthm}
\usepackage{array, boldline, makecell, booktabs}

\theoremstyle{definition}

\hypersetup{colorlinks=true, linkcolor=DarkRed, citecolor=DarkRed,urlcolor=Indigo,linktocpage}
\def\beq{\begin{eqnarray}}\def\eeq{\end{eqnarray}}
\def\be{\begin{equation}}\def\ee{\end{equation}}
\def\bs{\begin{split}}\def\es{\end{split}}

\usepackage{bm}

\def \zbr{\bar{z}}

\usepackage{mathrsfs}
\usepackage{slashed}
\newcommand{\scrip}{\mathscr{I}^{+}}
\newcommand{\scrim}{\mathscr{I}^{-}}

\begin{document}

\title{\bf Charges for Hypertranslations and Hyperrotations}

\author{\!\!\!\! Chethan Krishnan$^{a}$ and Jude Pereira$^{b}$\\ ~~~~\\
\it ${^a}$Centre for High Energy Physics,
\it Indian Institute of Science,\\ \it C.V. Raman Road, Bangalore 560012, India. \\
{\rm Email:} \textmd{chethan.krishnan.physics@gmail.com}\\
\it ${^b}$ Department of Physics, Arizona State University,\\ Tempe, Arizona 85287-1504, USA. \\
{\rm Email:} \textmd{jude.pereira@asu.edu}}

\begin{abstract}{Hypertranslations and hyperrotations are asymptotic symmetries of flat space, on top of the familiar supertranslations and superrotations. They were discovered in arXiv:2205.01422 by working in the Special Double Null (SDN) gauge, where $\scrip$ and $\scrim$ are approached along $null$ directions. It was observed there that while the hair degrees of freedom associated to these diffeomorphisms show up in the covariant surfaces charges, the diffeomorphisms themselves do not. This made their status intermediate in some ways between global symmetries and trivial gauge transformations, making interpretation ambiguous. In this paper, we revisit the fall-offs considered in arXiv:2205.01422 which were strictly subleading to Minkowski in conventional double null coordinates. We identify a new class of fall-offs where this assumption is relaxed, but whose charges 
nonetheless remain finite. Remarkably, the leading behavior is still Riemann flat, indicating that these are soft modes. With this more refined definition of asymptotic flatness, we show that leading hypertranslations and leading hyperrotations explicitly show up in the charges. This makes them genuine global symmetries of asymptotically flat Einstein gravity in the SDN gauge.  
We write down the new algebra of asymptotic Killing vectors that subsumes the BMS algebra.

}
\end{abstract}
\maketitle

{\noindent \bf Introduction}: When the cosmological constant $\Lambda$ is negative, it is widely believed that the radial direction  of the resulting anti-de Sitter (AdS) spacetime is holographically emergent \cite{Witten}. When $\Lambda$ is positive, less is known, but there are many suggestions in the literature that the timelike direction of de Sitter (dS) space may have a  holographic origin \cite{dS}. These observations make one suspect that it may be useful to view the holographic direction for flat space, which has $\Lambda=0$, as a null coordinate. Historically however, the null boundary of flat space is typically approached along a spacelike direction, eg., in the famous Bondi gauge \cite{Bondi}. 

With quite different motivations, various aspects of flat space were explored from a holographic perspective in \cite{Budhaditya, ACD, Jude}. Along the way, it was realized that a natural gauge for asymptotically flat space is the Special Double Null (SDN) gauge \cite{CJ1}, defined by
\begin{eqnarray}
g^{uu}=0,\ g^{vv}=0,\ g^{uA}=g^{vA}. \label{SDNbasic}
\end{eqnarray} 
Here $u$ and $v$ are null coordinates and $\scrip$ and $\scrim$ are covered (generically) by two separate patches around $v \rightarrow \infty$ and  $u \rightarrow -\infty$ respectively. The holographic directions are $v$ and $-u$ in these patches. 

The notion of a double null coordinate system has been explored in various contexts in the literature before, see eg. \cite{WaldText, Israel}. But usually in these settings, not enough constraints are imposed to fix all the coordinate freedom; they are therefore not genuine {\em gauge} choices. In fact in the context of mathematical relativity, the form of the double null metric that is sometimes written down (see eg., eqn (70) of \cite{Dafermos}) does {\em not} fall into the gauge we have presented above. This reflects a difference in philosophy. General relativists are interested in $\scrip$ as the eventual location of gravitational waves from localized objects. But if one is interested in graviton scattering, as perhaps necessary in quantum gravity, we need access to both $\scrip$ and $\scrim$. Our gauge has a natural $u \leftrightarrow -v$ symmetry which relates $\scrip$ and $\scrim$. This manifests an {\em asymptotic} CPT invariance \cite{CJ1}, which is believed to be a symmetry of quantum gravity in flat space \cite{Strominger}. 

In \cite{CJ2} we considered the most general asymptotic symmetry algebra in SDN gauge with fall-offs which are power laws in the respective null coordinate. Minkowski space in double null coordinates can be obtained by writing $u=t-r$ and $v=t+r$: 
\beq
ds^2=-du\ dv +2 \Big(\frac{v-u}{2}\Big)^2\gamma_{z\zbr}dz d\zbr \label{Mink}
\eeq
which has $g_{uA}=g_{vA}=0$. This suggested that one should allow fall-offs where $g_{uA}$ and $g_{vA}$ are at most $O(v^{-1})$ at $\scrip$. The result was found to contain new classes of non-trivial asymptotic diffeomorphisms on top of the BMS symmetries \cite{Barnich}. These were named $hypertranslations$, {\em subleading hypertranslations} and {\em subleading hyperrotations}\footnote{In \cite{CJ2}, the latter were simply called hyperrotations. But in the present paper we will find more leading counterparts to these AKVs which are more naturally called (leading) hyperrotations. Therefore the ones noted in \cite{CJ2} will be referred to as subleading hyperrotations in this paper.}. The algebra of asymptotic Killing vectors that extends the BMS algebra was identified and the covariant surface charges \cite{Iyer, Brandt} were computed. It was noted that these charges had non-trivial dependence on the corresponding ``hair'' (the metric parameters affected by these asymptotic diffeomorphisms). But at the same time, they did {\em not} contain the new asymptotic diffeomorphisms themselves, and therefore the interpretation of these charges was ambiguous. Typically for global symmetries that emerge from an asymptotic symmetry calculation, both the diffeomorphisms and the associated hair parameters appear in the charge expression. On the other hand, for trivial diffeomorphisms, neither the diffeomorphisms nor the parameters associated to them appear in the charges.  This made the status of hypertranslations and hyperrotations intermediate between global symmetries and trivial gauge transformations, making them challenging to interpret. Part of the problem here is that because we are working with null directions, the formalism that is most suited for our purposes is the covariant phase space approach of Wald and followers \cite{Iyer, Brandt}, while a more Hamiltonian-like formalism is perhaps more suited for interpretational purposes.  

In this paper, we will bypass this problem by identifying a new set of fall-offs which are {\em not} strictly subleading to \eqref{Mink}, but for which the charges are still finite. These fall-offs are presented in \eqref{boundarycondnew} and also in more detail in the Supplementary Material. In particular, our fall-offs will allow $g_{uA}=O(v^0)=g_{vA}$. A key feature of these fall-offs is that they can change the metric at an order more leading than \eqref{Mink}, and yet remarkably, we are able to show that their charges remain finite. 
In particular, a striking fact that we note 
is that demanding Riemann flatness $allows$ these terms.  This allows us to adopt the philosophy that there is nothing too sacred about the {\em specific form} in expression \eqref{Mink}, it is the demand of Riemann flatness that should be respected in deciding the leading behavior. We will find that Riemann flatness still leaves the possibility that these modes can be functions of the angular coordinates $(z, \zbr)$. We will eventually identify these as related to the hyperrotation hair. This should be compared to the familiar fact that purely angle dependent shear modes in Bondi gauge are soft hair associated to supertranslations, and turning them on can still leave the metric Riemann flat \cite{StromingerR}. Similarly in SDN gauge, turning on supertranslation hair or hypertranslation hair, leaves the metric Riemann flat. But both in Bondi gauge as well as in SDN gauge, the supertranslation and hypertranslation soft modes were subleading to the corresponding conventional form of the Minkowski metric. The new feature of hyperrotation hair here is that it is more leading than \eqref{Mink} while remaining Riemann flat. Riemann flatness in SDN gauge has many remarkable properties, which will be discussed in detail elsewhere \cite{CJRiemann}.


Once we adopt these relaxed fall-offs the nature of the calculation is parallel to that in \cite{CJ2}, even though technically more involved due to the increased number of metric functions that we start with. 
The result of this exercise is that we find that (a) the charges are still finite, (b) there is a new set of ${\rm Diff}(S^2)$ transformations (the leading hyperrotations) that appear before subleading hyperrotations but are subleading to superrotations, (c) both the hair parameters as well as the diffeomorphisms associated to the leading hypertranslations and leading hyperrotations appear in the charge expressions, on top of the conventional BMS quantities, (d) demanding Riemann flatness still allows soft hair associated to these  diffeomorphisms to appear in the metric, and (e) the algebra of the asymptotic symmetries is enhanced with respect to both the BMS algebra as well as the BBMS algebra of \cite{CJ2}. 

The next section contains the main results of this paper. To avoid repetition, we will only emphasize aspects of the discussion that are distinct from those in \cite{CJ2}. In particular, we will simply present the final algebra without presenting the details of the derivation -- the approach is identical to that in \cite{CJ2}, even though technically more involved.









{\noindent \bf{Results}}: We will work with SDN gauge discussed in \cite{CJ1}. The fall-offs are presented in great detail in the Supplementary Material in terms of functions appearing in the metric. Here we will write the fall-offs as 
\begin{subequations}\label{boundarycondnew}
    \begin{align}
        g^{uv} &= -2+O\big(v^{-1}\big)\\
        g^{AB} &= 4\gamma^{AB}\, v^{-2}+O\big(v^{-3}\big)\\
        g^{uA} &= g^{vA} = O\big(v^{-2}\big)
    \end{align}
\end{subequations}
Even though technically this is a small change from our previous paper, we emphasize that this is a pretty substantive departure from experience in other gauges. We are demanding that the metric be distinct from the conventional form Minkowski metric \eqref{Mink}, already at leading order. There are three reasons why we believe this is reasonable. Firstly, the charges remain finite. Secondly, demanding Riemann flatness does not force these terms to be zero. Thirdly, with this choice, we get a perfectly conventional structure for the leading hypertranslation and hyperrotation charges.

The asymptotic Killing vector conditions take the form:
\begin{subequations}\label{approxLie}
    \begin{align}
        \label{approxLieuv}
        \mathcal{L}_{\xi}g^{uv} &= O\big(v^{-1}\big)\\
        \label{approxLieuA}
        \mathcal{L}_{\xi}g^{uA} &= O\big(v^{-2}\big) \\
        \label{approxLievA}
        \mathcal{L}_{\xi}g^{vA} &= O\big(v^{-2}\big) \\
        \label{approxLieAB}
        \mathcal{L}_{\xi}g^{AB} &= O\big(v^{-3}\big)
    \end{align}
\end{subequations}
These and the exact Killing conditions \eqref{exactKilling}, lead to the solutions:
\begin{subequations}\label{finalxi}
    \begin{align}
        \xi^u &= f+ \frac{\xi^{u}_{(1)}}{v}+ \frac{\xi^{u}_{(2)}}{v^2}+ \frac{\xi^{u}_{(3)}}{v^3}
        +O\big(v^{-4}\big) \\
        \xi^v &= -\frac{\psi}{2}\, v+\xi^v_{(0)}+\frac{\xi^v_{(1)}}{v}+\frac{\xi^v_{(2)}}{v^2}+O\big(v^{-3}\big) \\
        \xi^A &= Y^A +\frac{\xi^A_{(1)}}{v}+\frac{\xi^A_{(2)}}{v^2}+\frac{\xi^A_{(3)}}{v^3}+O\big(v^{-4}\big)      
    \end{align}
\end{subequations}
where 
\begin{subequations}
    \begin{align}
        f &=\xi_{(0)}^{u}=\psi(z,\zbr)\, u/2 + T(z,\zbr),  \ \ {\rm with} \ \ \psi(z,\zbr)= D_AY^A \label{fdef}\\
        \xi^u_{(1)} &=\alpha^{A}_{2}\partial_Af \\
        \xi^u_{(2)} &=\frac{1}{2}\big(\alpha^{A}_{3}\partial_Af+\alpha^{A}_{2}\partial_A\xi^u_{(1)}\big)\\
        \xi^u_{(3)} &=\frac{1}{3}\big(\alpha^{A}_{4}\partial_Af+\alpha^{A}_{3}\partial_A\xi^u_{(1)}+\alpha^{A}_{2}\partial_A\xi^u_{(2)}\big)
    \end{align}
\end{subequations}
Here $T(z,\zbr)$ denotes supertranslations, and $Y^{z}(z), Y^{\zbr}(\zbr)$ denote superrotations. On top of the BMS diffeomorphisms, the $\xi^v_{(0)}, \xi^v_{(1)}$, $\xi^A_{(1)}$ and $\xi^A_{(2)}$ are also determined by the exact and asymptotic Killing conditions. The independent functions contained in them are \emph{hypertranslations} $\phi(z,\zbr)$,  \emph{sub-leading hypertranslations} $\tau(z,\zbr)$, {\em hyperrotations} $X^{A}(z,\zbr)$ and {\em sub-leading hyperrotations} $Z^{A}(z,\zbr)$ respectively. They are related to the $\xi^v_{(0)}, \xi^v_{(1)}, \xi^A_{(1)}, \xi^A_{(2)}$ via:
\begin{subequations}\label{xidef}
    \begin{align}
\xi^A_{(1)} &= X^{A} -2\, D^Af \\    
\xi^v_{(0)}&=\phi+ T+\triangle_\gamma T-\frac{1}{4}a^A_{2} D_A\psi-\frac{1}{2}D_AX^A \\
\xi^v_{(1)} &= \tilde\tau+\frac{1}{2}\mathscr{A}^A_2D_A\psi \\
\begin{split}
 \xi_{(2)}^A &= \tilde Z^{A}+\mathscr{C}^{AB}\, D_B\psi+\mathscr{A}^A_2\psi-u\, X^A+2\, u\,  D^A\xi^v_{(0)}\\&- u^2\, D^A\psi-\mathscr{L}_1\, D^A\psi \nonumber   
\end{split}
\end{align}
\end{subequations}
We have introduced $\tilde\tau$ and $\tilde Z^{A}$ for convenience which are related to the sub-leading hypertranslations $\tau$ and sub-leading hyperrotations $Z^{A}$ via
\begin{equation}
    \begin{aligned}
    \tilde{\tau} &=\tau-\frac{1}{4}\, a_3^A\, D_A\psi \\&+\big(D_{\zbr}c^{z\zbr}- D_{z}c^{zz}+\gamma^{z\zbr}D_zD_{\zbr}a^z_2- \gamma^{z\zbr}D^2_{\zbr}a^{\zbr}_2\big)\, D_zT \\&+\big(D_{z}c^{z\zbr}- D_{\zbr}c^{\zbr\zbr}+\gamma^{z\zbr}D_{\zbr}D_z a^{\zbr}_2-\gamma^{z\zbr}D^2_z a^z_2\big)\, D_{\zbr}T \\&+a^A_2D_A\xi^v_{(0)}+a^A_2D_AT +\frac{1}{4}a^A_2D_A\big(a^B_2D_B\psi\big)
    \end{aligned}
\end{equation}
\begin{equation}\label{Z1}
    \begin{aligned}
    \tilde Z^z &=Z^z+c^{zz}\, D_zT+c^{z\zbr}\, D_{\zbr}T+T D_z c^{zz}-T D_{\zbr}c^{z\zbr}\\&+a^z_2\, \xi^v_{(0)}-\frac{1}{2}X^BD_Ba^z_2+\frac{1}{2}a^B_2D_BX^z-\gamma^{z\zbr}D_{\zbr}a^{\zbr}_2D_{\zbr}T\\&-\gamma^{z\zbr}D_{\zbr}a^z_2D_zT+\frac{1}{4}a^z_2a^{\zbr}_2D_{\zbr}\psi-\gamma^{z\zbr}a^{\zbr}_2D^2_{\zbr}T+\frac{1}{4}\big(a^z_2\big)^2D_z\psi\\&-\frac{1}{2}a^z_2\Delta_{\gamma}T-\gamma^{z\zbr}TD_{z}D_{\zbr}a^z_2+\gamma^{z\zbr}TD^2_{\zbr}a^{\zbr}_2
    \end{aligned}
\end{equation}
\begin{equation}\label{Z2}
    \begin{aligned}
    \tilde Z^{\zbr} &=Z^{\zbr}+c^{\zbr\zbr}\, D_{\zbr}T+c^{z\zbr}\, D_{z}T+T D_{\zbr}c^{\zbr\zbr}-T D_{z}c^{z\zbr}\\&+a^{\zbr}_2\, \xi^v_{(0)}-\frac{1}{2}X^BD_B a^{\zbr}_2+\frac{1}{2}a^B_2D_BX^{\zbr}-\gamma^{z\zbr}D_{z}a^{z}_2D_{z}T\\&-\gamma^{z\zbr}D_{z}a^{\zbr}_2D_{\zbr}T+\frac{1}{4}a^{\zbr}_2 a^{z}_2D_{z}\psi-\gamma^{z\zbr}a^{z}_2D^2_{z}T+\frac{1}{4}\big(a^{\zbr}_2\big)^2D_{\zbr}\psi\\&-\frac{1}{2}a^{\zbr}_2\Delta_{\gamma}T-\gamma^{z\zbr}TD_{\zbr}D_{z}a^{\zbr}_2+\gamma^{z\zbr}TD^2_{z}a^{z}_2
    \end{aligned}
\end{equation}

These expressions are significantly more complicated than those in \cite{CJ2}, so let us pause to explain some of the details.
The integration ``constants'' in the shear are\footnote{The notation here is slightly different from that in \cite{CJ2}.} introduced via
\beq \mathcal{C}_{AB}(u,z,\zbr)=c_{AB}(z,\zbr)+\int_{-\infty}^{u}du'\mathcal{N}_{AB}(u',z,\zbr)
\eeq
with $\mathcal{N}_{AB}\equiv \partial_u\mathcal{C}_{AB}$, being the SDN news tensor. Similarly, we have defined the integration ``constant'' in $\alpha^A_2$ as
\beq \alpha^A_2(u,z,\zbr)=a^A_2(z,\zbr)+\int_{-\infty}^{u}du'\beta^A_2(u',z,\zbr)\eeq where $\beta^A_2 \equiv \partial_u\alpha^A_2$.
See \cite{CJ1, CJ} for a discussion on integrals of this type that are defined from $\scrip_-$ to $u$. On shell (ie., when Einstein equations hold), we have $\mathcal{N}_{z\zbr}=0$  and $\beta^A_2=0$, so we will have
\beq\begin{aligned}
    \mathcal{C}^{z\zbr}(u,z,\zbr) &=c^{z\zbr}(z,\zbr)\\
    \alpha^A_2(u,z,\zbr) &=a^A_2(z,\zbr) \label{onshell}
\end{aligned}
\eeq
In addition to this, the Einstein constraints also require that $\lambda_1=0$.
For $\xi_{(2)}^A$, combining all the relevant equations, we can write \cite{CJ} 
\beq\label{xi2A}\begin{aligned}
    \partial_u\xi_{(2)}^A &= \mathcal{C}^{AB}\, D_B\psi-2\, u\, D^A\psi+2\, D^A\xi^v_{(0)} +\alpha^A_2\psi \\&-\lambda_1D^A\psi-X^A \\ \implies \xi_{(2)}^A &= \mathscr{C}^{AB}\, D_B\psi - u^2\, D^A\psi +2\, u\, D^A\xi^v_{(0)}+\mathscr{A}^A_2\psi \\&-\mathscr{L}_1D^A\psi-u\, X^A +\tilde Z^{A}(z,\zbr)
\end{aligned}\eeq
The $u$-independence of $\psi$, $\xi^v_{(0)}$ and $X^A$ has been used in writing the integrated version in the second step. Also $\mathscr{C}^{AB}$, $\mathscr{A}^A_2$ and $\mathscr{L}_2$ have been defined via
\begin{eqnarray}
\partial_u\mathscr{C}^{AB} & =& \mathcal{C}^{AB}(u,z,\zbr)\\
\partial_u\mathscr{A}^{A}_2 &= &\alpha^A_2(u,z,\zbr)\\
\partial_u\mathscr{L}_1 & =& \lambda_1(u,z,\zbr)
\end{eqnarray}
As in \cite{CJ2}, $\tilde Z^{A}(z,\zbr)$ is taken as the $u$-independent piece in $\xi^A_{(2)}$. The shift is done on $\tilde Z^A$ via \eqref{Z1}-\eqref{Z2} and the result is what we call sub-leading hyperrotations $Z^A$.

The rest of the notation follows that of \cite{CJ2}. As emphasized there, the idea in \eqref{xidef} is to do certain shifts so that the structure of the diffeomorphisms is cleanest. This ``diagonalizes'' the algebra of diffeomorphisms. The philosophy here is identical, even though the expressions are more complicated.

The hair associated to the various diffeomorphisms are therefore as follows: supertranslations $T(z,\zbr)$ are associated to the $u$-independent shifts in $\mathcal{C}_{zz}$ and $\mathcal{C}_{\zbr\zbr}$, hypertranslations $\phi(z,\zbr)$ are associated to the $u$-independent shifts of $\mathcal{C}_{z\zbr}$, subleading hypertranslations $\tau(z,\zbr)$ are associated to $u$-independent shifts of $\lambda_2$, hyperrotations $X^{A}(z,\zbr)$ are associated to $u$-independent shifts of $\alpha_{2}^{\ A}$ and sub-leading hyperrotations $Z^{A}(z,\zbr)$ are associated to $u$-independent shifts of $\alpha_{3}^{\ A}$. As in Bondi gauge, we also have superrotations $Y^z(z), Y^{\zbr}(\zbr)$. 
The shifts involved in the definitions of $\phi, \tau$, $X^A$ and $Z^A$ are detailed in the Supplementary Material.   
As in \cite{CJ2}, supertranslations and leading-\&-subleading hypertranslations are diffeomorphisms of $u$ and $v$ respectively. Leading hyperrotations were not present in \cite{CJ2}, but both leading and subleading hyperrotations  are subleading to superrotations on the sphere. 

{
We will define the ``Beyond BBMS'' algebra $\mathfrak{b}^2$-$\mathfrak{bms}_4$ as the asymptotic symmetry algebra of the nine non-trivial diffeomorphisms --  supertranslations, superrotations, hypertranslations \& subleading hypertranslations, and hyperrotations \& subleading hyperrotations. 
Following \cite{Barnich, CJ2}, we define the bracket
\begin{widetext}
\begin{eqnarray}
\label{YTtaubracket}\big(\widehat{Y},\widehat{T},\widehat{\phi},\widehat{\tau},\widehat{X},\widehat{Z}\big)=\big[(Y_1,T_1,
\phi_1,\tau_1,X_1,Z_1),(Y_2,T_2,\tau_2,\phi_2,X_2,Z_2)\big]
\end{eqnarray}
\end{widetext}
The notation is the natural generalization of that in \cite{Barnich, CJ2} and the reader should consult those papers for the detailed definitions.  
The new algebra is defined via $\widehat{Y}$, $\widehat{T}$, $\widehat{\phi}$, $\widehat{\tau}$, $\widehat{X}$ and $\widehat{Z}$ given by the following expressions:
\begin{widetext}
\begin{subequations}\label{allhats}
    \begin{align}\label{phihat}
    \widehat{Y}^A &= Y_1^B\, \partial_B Y_2^A-Y_2^B\, \partial_B Y_1^A \\
        \widehat{T} &= Y_1^A\, \partial_A T_2-Y_2^A\, \partial_A T_1+\frac{1}{2}\, (T_1\, \psi_2-T_2\, \psi_1). \\
         \begin{split}\label{YhatThat}
    \widehat{\phi} &=\frac{1}{2}(\psi_1\phi_2-\psi_2\phi_1)+\big(Y_{1}^{A}\partial_A\phi_2-Y_{2}^{A}\partial_A\phi_1\big)
    \end{split}\\
    \begin{split}\label{tauhat}
    \widehat{\tau} &= (\psi_1\tau_2-\psi_2\tau_1)+\big(Y_1^A\partial_A\tau_2-Y_2^A\partial_A\tau_1\big)
    \end{split}\\
    \begin{split}\label{Xhat}
    \widehat{X}^A &= \frac{1}{2}\big(\psi_{1}X^A_{2}-\psi_{2}X^A_{1}\big)+\big(Y_1^{B}\partial_B X^A_2-Y_2^{B}\partial_B X^A_1\big)+\big(X^B_1\partial_BY_2^{A}-X^B_2\partial_BY_1^{A}\big)
    \end{split}\\
    \begin{split}\label{Zhat}
    \widehat{Z}^A &= \big(\psi_{1}Z^A_{2}-\psi_{2}Z^A_{1}\big)+\big(Y_1^{B}\partial_B Z^A_2-Y_2^{B}\partial_B Z^A_1\big)+\big(Z^B_1\partial_BY_2^{A}-Z^B_2\partial_BY_1^{A}\big)
    \end{split}
    \end{align}
\end{subequations}
\end{widetext}
This is what we call the $\mathfrak{b}^2$-$\mathfrak{bms}_4$ algebra. The fact that these nine non-trivial diffeomorphisms form a closed algebra is checked by the same procedure as outlined in \cite{CJ2}. The calculations are straightforward but lengthier variations of those there. In order to identify the capped quantities, we need to consider the Barnich-Troessaert bracket $[\xi_1,\xi_2]_M$ of two AKVs $\xi_1$ and $\xi_2$ \cite{Barnich, CJ2}. The structure is parallel to that presented in \cite{CJ2}, with a notable difference in the $A$-component which takes the form
\begin{equation}\label{BBMSapprox}
        [\xi_1,\xi_2]^A_M = \widehat{Y}^A +\frac{\widehat{\xi}^A_{(1)}}{v}+\frac{\widehat{\xi}^A_{(2)}}{v^2}+O\big(v^{-3}\big).      
\end{equation}
In computing all four components of the Barnich-Troessaert bracket, we need $\widehat{\xi}^v_{(0)}, \widehat{\xi}^v_{(1)}$, $\widehat{\xi}_{(1)}^A$ and $\widehat{\xi}_{(2)}^A$, which are defined as in \eqref{xidef} but with $Y^A, T, \phi, \tau, X^A, Z^A$ replaced by their capped versions, defined in  \eqref{allhats}. 

Equations \eqref{allhats} define the $\mathfrak{b}^2$-$\mathfrak{bms}_4$ algebra.  Setting the hyperrotations $X^A$ to zero results in the BBMS algebra of \cite{CJ2}, and setting $\phi, \tau$ and $Z^A$ as well to zero results in the familiar BMS algebra \cite{Barnich}.

{\noindent \bf{Discussion}}: In this paper, we observed that demanding finite covariant surface charges in Einstein gravity allows fall-offs that are not necessarily subleading to \eqref{Mink}. Turning on the soft modes associated to supertranslations and leading hypertranslations/hyperrotations takes us beyond \eqref{Mink} even though the metric is still Riemann flat. We exploited this fact to work with fall-offs that allowed these modes, to show that the covariant surface charges contain these diffeomorphisms as well as the associated soft hair. This places them on an equal footing with conventional global symmetries (eg. supertranslations), resolving some of the ambiguities pointed out in \cite{CJ2}.

Of course, these results open up further questions. Our work strongly suggests that the charges associated to hypertranslations should be interpreted as soft, so it would be interesting to connect these results to soft theorems (perhaps to the subsubleading soft graviton theorem of \cite{Cachazo}?) and also to new memory effects. Some of these questions are currently under investigation. Hypertranslations have many similarities to supertranslations, but there are also crucial distinctions. The lowest modes of supertranslations are simply the action of Poincare translations on the boundary $(u, z, \zbr)$. Hypertranslations on the other hand are truly distinct from bulk translations -- we have already subtracted out the supertranslations in our shifted diffeomorphisms, when defining hypertranslations. It should be clear from \eqref{finalxi} that the interpretation of hypertranslations is more like a {\em bulk} diffeomorphism at infinity (note that infinity is along the null direction $v$ in SDN gauge). It is more naturally compared to $\xi^r$ than $\xi^u$ in Bondi gauge. 

A related  interesting feature of hypertranslations and their associated hair is that they can be spherically symmetric. This raises a subtlety in the usual statement of Birkhoff's theorem, which will be discussed in an upcoming work. Note that while supertranslations allow soft hair on Schwarzschild, the only spherically symmetric supertranslation is a time translation, so this subtlety does not arise for Schwarzschild in Bondi gauge. It is also important to emphasize that hypertranslations should be distinguished from the shifts in $v$ at the {\em past} boundary $\scrim$. The latter are simply supertranslations, but now acting in the past. What we mean by hypertranslations are shifts in $v$ at $\scrip$. There is no obvious connection between the two (other than the future-past matching at $i^0$ that was discussed in \cite{CJ1}) because these coordinates live in different charts. 



What about subleading hypertranslations and subleading hyperrotations? They do not show up in the charges even with the new fall-offs, but their associated hair was present both in \cite{CJ2} as well as here. So their interpretation remains ambiguous. It is natural to consider the sub-algebra obtained by setting the subleading hypertranslations/hyperrotations to zero. This would mean that we are working with supertranslations,  superrotations, leading hypertranslations and leading hyperrotations. This is a natural generalization of the conventional BMS algebra in the SDN gauge; it is clearly of interest to study it more closely. One could also consider the even simpler generalization of BMS, obtained by adding only the leading hypertranslations and suppressing the leading hyperrotations. This algebra has the advantage that we are not turning on diffeomorphisms on the sphere, but only the Virasoro (super)rotations. While it may be difficult to conclusively argue for such a choice from a purely asymptotic symmetry perspective, it is natural from a celestial holography perspective \cite{Schwarz}. This is the algebra of supertranslations, (leading) hypertranslations and superrotations.

\section*{Acknowledgments}
 
We thank Sudip Ghosh and Sarthak Talukdar for discussions.

\vspace{0.3in} 
\onecolumngrid
{\begin{center}\bf \Large{Supplementary material}\end{center} }

\section{Refined Fall-Offs}

In this section, we will present the falloffs in some detail. Our emphasis will be on the distinctions from those presented in \cite{CJ2}. We start with a quick review of the notation: in $d+1$ dimensions, the SDN gauge \cite{CJ1} is defined by eqn \eqref{SDNbasic}. We will restrict ourselves to 3+1 dimensions here. The exact Killing vector equations are
\begin{eqnarray}\label{exactKilling}
        \mathcal{L}_{\xi} g^{uu} = 0, \
        \mathcal{L}_{\xi} g^{vv} = 0, \ 
        \mathcal{L}_{\xi} g^{uA} = \mathcal{L}_{\xi} g^{vA}
\end{eqnarray}
and we will write the general metric in this gauge as
\beq \label{doublenull}
ds^2=-e^{\lambda}du\ dv +\Big(\frac{v-u}{2}\Big)^2\Omega_{AB}(dx^A-\alpha^A du-\alpha^A dv)(dx^B-\alpha^B du-\alpha^B dv)
\eeq 
In \cite{CJ2}, we presented a set of fall-offs in terms of the functions in this ansatz, which the reader should consult. The fall-offs we consider in this paper are distinct in the following functions:
\begin{subequations}\label{polyfalloffs}
    \begin{align}
        \label{polylambda}
        \lambda(u,v,z,\zbr) &= \frac{\lambda_{1}(u,z,\zbr)}{v}+\frac{\lambda_{2}(u,z,\zbr)}{v^2}+\frac{\lambda_{3}(u,z,\zbr)}{v^3}+\frac{\lambda_{4}(u,z,\zbr)}{v^4}+O\big(v^{-5}\big) \\
        \label{polyalphaz}
        \alpha^z(u,v,z,\zbr) &= \frac{\alpha^{z}_{\ 2}(u,z,\zbr)}{v^2}+\frac{\alpha^{z}_{\ 3}(u,z,\zbr)}{v^3}+\frac{\alpha^{z}_{\ 4}(u,z,\zbr)}{v^4}+\frac{\alpha^{z}_{\ 5}(u,z,\zbr)}{v^5}+O\big(v^{-6}\big)\\
        \label{polyalphaw}
        \alpha^{\zbr}(u,v,z,\zbr) &= \frac{\alpha^{\zbr}_{\ 2}(u,z,\zbr)}{v^2}+\frac{\alpha^{\zbr}_{\ 3}(u,z,\zbr)}{v^3}+\frac{\alpha^{\zbr}_{\ 4}(u,z,\zbr)}{v^4}+\frac{\alpha^{\zbr}_{\ 5}(u,z,\zbr)}{v^5}+O\big(v^{-6}\big)
    \end{align}
\end{subequations}
In terms of the metric, this results in the fall-offs:
\begin{subequations}\label{boundarycond}
    \begin{align}
        g_{uu} &= g_{vv}= O\big(v^{-2}\big)\\
        g_{uv} &= -\frac{1}{2}+O\big(v^{-1}\big)\\
        g_{AB} &= \frac{1}{4}\, \gamma_{AB}\, v^2+O(v)\\
        g_{uA} &= g_{vA} = O\big(v^0\big)
    \end{align}
\end{subequations}
Compared to the discussion in \cite{CJ2}, we also allow $\alpha^A_2$ as the $O(1/v^2)$ term in the $\alpha^A$ fall-off. Just like $C_{z\zbr}$, $\alpha^A_2$ also turns out to be $u$-independent once we demand Einstein equations. Hence it is an integration ``constant'' in  Einstein constraints in the language of \cite{CJ1, CJ, CJ2}.  

Demanding Ricci (or Riemann) flatness forces $\lambda_1$ to be zero and $\alpha_2^A$ to be functions only of the angles. We have kept them general in the discussions of the AKVs because they can be defined on arbitrary backgrounds, without worrying about the equations satisfied by those backgrounds. But one can in principle start {\em a-priori }with fall-offs \eqref{polyfalloffs} where $\lambda_1$ is set to zero and $\alpha_2^A$ are functions only of $z$ and $\zbr$. Some of the expressions we have presented will simplify somewhat in that case, but the main results do not change.


\section{Diffeomorphism Shifts}



As in \cite{CJ2} we will define the various diffeomorphisms after a suitable shift in the fall-off coefficient of $\xi$. This is more elaborate in the present case, and we discuss them in detail below. The philosophy behind these shifts was discussed in \cite{CJ2}.

{\bf Hyperrotations:} The  simplest case arises for the leading hyperrotations $X^A(z,\zbr)$, so we start with them. From the exact Lie derivative conditions, we obtain the following constraint on $\xi^{A}_1$,
\begin{equation}
    \partial_u\xi^A_1 = -D^A\psi
\end{equation}
which on integrating both sides becomes
\begin{equation}\label{xiA1}
    \xi^A_1 = \tilde{X}^A-u\, D^A\psi
\end{equation}
The metric function corresponding to leading hyperrotations is $\alpha^A_2$. Under the action of AKVs, the transformation of $\alpha^A_2$ can be obtained by evaluating $\delta_{\xi}g^{uA}=\mathcal{L}_{\xi}g^{uA}$ at $O(v^{-2})$ as follows
\begin{equation}\label{deltaalphaA2}
\delta\alpha^A_2 = \Big[f\partial_u+\mathcal{L}_Y+\frac{\psi}{2}\Big]\alpha^A_2+\tilde{X}^A -u\, D^A\psi +2\, D^Af
\end{equation}
where
\beq\mathcal{L}_Y\alpha_{2}^{A} = Y^B\, \partial_B \alpha_{2}^{A}- \alpha_{2}^{B}\, \partial_BY^A.\eeq
is the Lie derivative of $\alpha^A_2$ with respect to $Y^A$. Recalling that on-shell $\alpha^A_2 =a^A_2(z,\zbr)$ and substituting $f =\psi(z,\zbr)\, u/2 + T(z,\zbr)$, we obtain
\begin{equation}
\delta a^A_2 = \Big[\mathcal{L}_Y+\frac{\psi}{2}\Big]a^A_2+\tilde{X}^A+2\, D^AT
\end{equation}
Next we would like to interpret $X^A(z,\zbr)$ as the diffeomorphism that causes $\alpha^A_2$ to be turned on if it was initially zero. This immediately suggests the following shift
\begin{equation}
    \tilde{X}^A = X^A-2\, D^AT
\end{equation}
Substituting this in \eqref{xiA1} and using $f=\big(\psi/2\big)u+T$ yields
\begin{equation}
    \xi^A_1 = X^A-2\, D^Af
\end{equation}

{\bf Hypertranslations:} In the case of the leading hypertranslations $\phi(z,\zbr)$, the shift is of the form
\begin{equation}\label{phishift}
    \xi^v_{(0)}=\phi+ T+\triangle_\gamma T-\frac{1}{4}a^A_2 D_A\psi-\frac{1}{2}D_AX^A
\end{equation}
This reduces to the form presented in \cite{CJ2} when the hyperrotations and their hair are set to zero. 
The change in $\mathcal{C}_{z\zbr}$ can be computed by evaluating $\delta_{\xi}g^{z\zbr}=\mathcal{L}_{\xi}g^{z\zbr}$ at $O(v^{-3})$. The result is 
\beq\label{deltaCzw} \delta\mathcal{C}_{z\zbr} = \Big[f\, \partial_u+\mathcal{L}_{Y}-\frac{1}{2}\, \psi\Big]\, \mathcal{C}_{z\zbr}-4\, \partial_z\partial_{\zbr}f+2\, \gamma_{z\zbr}\Big(\xi^v_{(0)}-f-\frac{u}{2}\, \psi+\frac{1}{2}D_AX^A+\frac{1}{4}\alpha^A_2D_A\psi\Big) \eeq
Here $\mathcal{L}_Y$ is the Lie derivative of $\mathcal{C}_{z\zbr}$ with respect to $Y^A$ defined as in \cite{CJ2}:
\beq\mathcal{L}_{Y}\mathcal{C}_{z\zbr}=Y^A\partial_A\mathcal{C}_{z\zbr}+\big(\partial_AY^A\big)\mathcal{C}_{z\zbr}\eeq
On-shell we have $\mathcal{C}_{z\zbr}=c_{z\zbr}(z,\zbr)$ and $\alpha^A_2 =a^A_2(z,\zbr)$. Using these and substituting $f =\psi(z,\zbr)\, u/2 + T(z,\zbr)$, we obtain
\beq\delta c_{z\zbr}=\Big[\mathcal{L}_{Y}-\frac{1}{2}\, \psi\Big]c_{z\zbr}+2\gamma_{z\zbr}\Big(\xi^v_{(0)}-T-\Delta_{\gamma}T+\frac{1}{2}D_AX^A+\frac{1}{4}a^A_2D_A\psi\Big) \label{goldstone}\eeq

It is clear that $\xi^v_{(0)}$ mixes with supertranslations, superrotations and leading hyperrotations. We wish to remove this mixing, so that we can interpret $\phi(z,\zbr)$ as the diffeomorphism that causes $c_{z\zbr}$ to be turned on if it was initially zero. 
From this it follows that the shift is $ \xi^v_{(0)}=\phi+ T+\triangle_\gamma T-\frac{1}{4}a^A_2 D_A\psi-\frac{1}{2}D_AX^A$, as we presented above. This 
defines hypertranslations, $\phi(z,\zbr)$. Note that in deriving the algebra for hypertranslations, we have made use of the identity
\begin{equation}
    \delta_{\xi}\xi^v_{(0)}=-\frac{1}{4}\big(\delta a^A_2\big)D_A\psi
\end{equation}
where we have demanded that $\delta_{\xi}\phi=0$ and $\delta_{\xi}X^A=0$.
This shifted definition above of the hypertranslations ensures the vanishing of the hatted $\widehat \phi$ on the left hand side of algebra, when $\phi_1$ and $\phi_2$ are zero. As we pointed out in \cite{CJ2}, this feature can be viewed as one of the motivations behind doing the shifts. This generalizes to the other diffeomorphisms as well.



 
{\bf Subleading Hyperrotations:} Now we turn to the case of subleading hyperrotations $Z^A(z,\zbr)$ and the corresponding metric functions $\alpha^A_3$. 
The same procedure as in \cite{CJ2} now yields
\begin{equation}
    \begin{aligned}
    \label{deltaalpha3A}\delta\alpha_{3}^{z} &= \big[f\, \partial_u+\mathcal{L}_{Y}+\psi\big]\alpha_3^{z}+2\, \xi_{(2)}^{z}+4\, u\, D^{z}f-2\, \mathcal{C}^{zB}\, D_{B}f-2\, \alpha^z_2\, \xi^v_{(0)}+X^BD_B\alpha^z_2-\alpha^B_2D_BX^z \\& +2\, \gamma^{z\zbr}D_{\zbr}\alpha^{\zbr}_2D_{\zbr}f+2\, \gamma^{z\zbr} D_{\zbr}\alpha^z_2D_zT-\frac{1}{2}\alpha^z_2\alpha^{\zbr}_2D_{\zbr}\psi+2\, \gamma^{z\zbr}\alpha^{\zbr}_2D_{\zbr}^2T+2u\, \gamma^{z\zbr}\alpha^{\zbr}_2D_{\zbr}^2\psi\\& -u\, \gamma^{z\zbr}D_{z}\alpha^z_2D_{\zbr}\psi-\frac{1}{2}\big(\alpha^z_2\big)^2D_z\psi-2u\, \alpha^z_2\psi+\alpha^z_2\Delta_{\gamma}T + 2\, \lambda_1 D^zf + \partial_u\alpha^z_2\big(\alpha^A_2\, D_Af\big)
    \end{aligned}
\end{equation}
where the Lie derivative is defined as
\beq\mathcal{L}_Y\alpha_{3}^{A} = Y^B\, \partial_B \alpha_{3}^{A}- \alpha_{3}^{B}\, \partial_BY^A.\eeq
Note that in obtaining the above equation, we have used \eqref{deltaalphaA2} along with 
\beq\delta\lambda_1 = \Big[f\, \partial_u+\mathcal{L}_Y+\frac{1}{2}\psi\Big]\lambda_1+\partial_u\alpha^A_2\, D_Af\eeq
which has been obtained by evaluating $\delta_{\xi}g^{uv}=\mathcal{L}_{\xi}g^{uv}$ at $O(v^{-1})$ where $\mathcal{L}_{Y}\lambda_1=Y^A\partial_A\lambda_1$ is the Lie derivative of $\lambda_1$ with respect to $Y^A$. 
On-shell, we have \cite{CJ}
\beq\label{alpha3z}\begin{aligned}\partial_u\alpha_{3}^{z} &= -2\, D_{z}\mathcal{C}^{zz}+2\, D_{\zbr}c^{z\zbr}+2\, \gamma^{z\zbr}D_zD_{\zbr}a^z_2-2\, \gamma^{z\zbr}D^2_{\zbr}a^{\zbr}_2 \\ \implies \alpha^z_{3}(u,z,\zbr) &= -2\, D_{z}\mathscr{C}^{zz}+u\big(2\, D_{\zbr}c^{z\zbr}+2\, \gamma^{z\zbr}D_zD_{\zbr}a^z_2-2\, \gamma^{z\zbr}D^2_{\zbr}a^{\zbr}_2\big)+a_3^z(z,\zbr)\end{aligned}\eeq
and a similar equation for $\alpha^{\zbr}_{3}(u,z,\zbr)$.
Recalling that on-shell $\lambda_1=0$, substituting \eqref{onshell}, \eqref{alpha3z}, \eqref{xi2A} and \eqref{fdef} into \eqref{deltaalpha3A} and extracting the $u$-independent terms, we find
\begin{equation}
    \begin{aligned}
    \label{deltaaz}\delta a_3^z &=\big[\mathcal{L}_Y+\psi\big]a_3^z+2\, \tilde Z^z-2\, c^{zz}\, D_zT-2\, c^{z\zbr}\, D_{\zbr}T-2\, T\, D_zc^{zz}+2\, T D_{\zbr}c^{z\zbr}-2\, a^z_2\, \xi^v_{(0)}+X^B\, D_Ba^z_2\\&-a^B_2\, D_BX^z+2\, \gamma^{z\zbr}D_{\zbr}a^{\zbr}_2\, D_{\zbr}T+2\, \gamma^{z\zbr}D_{\zbr}a^z_2\, D_zT-\frac{1}{2}a^z_2 a^{\zbr}_2\, D_{\zbr}\psi+2\, \gamma^{z\zbr}a^{\zbr}_2\, D^2_{\zbr}T-\frac{1}{2}\big(a^z_2\big)^2\, D_z\psi\\&+a^z_2\, \Delta_{\gamma}T+2\, \gamma^{z\zbr}T\, D_{z}D_{\zbr}a^z_2-2\, \gamma^{z\zbr}T\, D^2_{\zbr}a^{\zbr}_2
    \end{aligned}
\end{equation}
As in \cite{CJ2}, the inhomogeneous part of the variation gives the shift: 
\begin{equation}\label{Zzshift}
    \begin{aligned}
    \tilde Z^z &=Z^z+c^{zz}\, D_zT+c^{z\zbr}\, D_{\zbr}T+T D_z c^{zz}-T D_{\zbr}c^{z\zbr}+a^z_2\, \xi^v_{(0)}-\frac{1}{2}X^BD_Ba^z_2 \\&+\frac{1}{2}a^B_2D_BX^z-\gamma^{z\zbr}D_{\zbr}a^{\zbr}_2D_{\zbr}T-\gamma^{z\zbr}D_{\zbr}a^z_2D_zT+\frac{1}{4}a^z_2a^{\zbr}_2D_{\zbr}\psi-\gamma^{z\zbr}a^{\zbr}_2D^2_{\zbr}T+\frac{1}{4}\big(a^z_2\big)^2D_z\psi\\&-\frac{1}{2}a^z_2\Delta_{\gamma}T-\gamma^{z\zbr}TD_{z}D_{\zbr}a^z_2+\gamma^{z\zbr}TD^2_{\zbr}a^{\zbr}_2
    \end{aligned}
\end{equation}
For completeness, we also present the result for $\alpha^{\zbr}_{3}(u,z,\zbr)$, which gives an analogous shift for the $\zbr$-component of the subleading hyperrotations: 
\begin{equation}\label{Zwshift}
    \begin{aligned}
    \tilde Z^{\zbr} &=Z^{\zbr}+c^{\zbr\zbr}\, D_{\zbr}T+c^{z\zbr}\, D_{z}T+T D_{\zbr}c^{\zbr\zbr}-T D_{z}c^{z\zbr}+a^{\zbr}_2\, \xi^v_{(0)}-\frac{1}{2}X^BD_B a^{\zbr}_2\\&+\frac{1}{2}a^B_2D_BX^{\zbr}-\gamma^{z\zbr}D_{z}a^{z}_2D_{z}T-\gamma^{z\zbr}D_{z}a^{\zbr}_2D_{\zbr}T+\frac{1}{4}a^{\zbr}_2 a^{z}_2D_{z}\psi-\gamma^{z\zbr}a^{z}_2D^2_{z}T+\frac{1}{4}\big(a^{\zbr}_2\big)^2D_{\zbr}\psi\\&-\frac{1}{2}a^{\zbr}_2\Delta_{\gamma}T-\gamma^{z\zbr}TD_{\zbr}D_{z}a^{\zbr}_2+\gamma^{z\zbr}TD^2_{z}a^{z}_2
    \end{aligned}
\end{equation}

The point of the shifts is that after doing them, the $Z^A$'s are the independent diffeomorphisms. So it is natural to demand  
\beq
\delta_\xi Z^A =0.
\eeq
This leads to 
\begin{equation}\label{deltaZz}
    \begin{aligned}
    \delta_{\xi}\tilde Z^z &=\big(\delta c^{zz}\big)\, D_zT+\big(\delta c^{z\zbr}\big)\, D_{\zbr}T+T \big(D_z\delta c^{zz}\big)-T \big(D_{\zbr}\delta c^{z\zbr}\big)+\big(\delta a^z_2\big)\, \xi^v_{(0)}+a^z_2\, \big(\delta_{\xi}\xi^v_{(0)}\big)-\frac{1}{2}X^B\big(D_B\delta a^z_2\big)\\&+\frac{1}{2}\big(\delta a^B_2\big)D_BX^z-\gamma^{z\zbr}\big(D_{\zbr}\delta a^{\zbr}_2\big)D_{\zbr}T-\gamma^{z\zbr}\big(D_{\zbr}\delta a^z_2\big)D_zT+\frac{1}{4} a^{\zbr}_2\big(\delta a^z_2\big)D_{\zbr}\psi+\frac{1}{4} a^z_2\big(\delta a^{\zbr}_2\big)D_{\zbr}\psi-\gamma^{z\zbr}\big(\delta a^{\zbr}_2\big)D^2_{\zbr}T\\&+\frac{1}{2} a^z_2\big(\delta a^z_2\big)D_z\psi-\frac{1}{2}\big(\delta a^z_2\big)\Delta_{\gamma}T-\gamma^{z\zbr}T\big(D_{z}D_{\zbr}\delta a^z_2\big)+\gamma^{z\zbr}T\big(D^2_{\zbr}\delta a^{\zbr}_2\big)
    \end{aligned}
\end{equation}
with a similar expression for $\delta_{\xi}\tilde Z^{\zbr}$. 
When computing the algebra for the shifted subleading hyperrotations $Z^A$, these expressions come in handy for cancelling out certain unpleasant pieces, and leading to 
the simple form of our final algebra \eqref{allhats}. 

{\bf Subleading Hypertranslations:} Following the same procedure as in \cite{CJ2}, we find \begin{equation}\begin{aligned}
    \delta\lambda_2 &=\big[f\partial_u+\mathcal{L}_Y+\psi\big]\lambda_2-\frac{1}{4}\, \alpha^A_3\, D_A\psi+\frac{1}{2}\, \partial_u\alpha^A_3\, D_Af-\xi^v_{(1)}+\alpha^A_2D_A\xi^v_{(0)}+\alpha^A_2D_AT +\frac{1}{4}\alpha^A_2D_A\big(\alpha^B_2D_B\psi\big) \\&-\lambda_1\, \xi^v_{(0)} +\xi^A_{(1)}D_A\lambda_1 +\partial_u\lambda_1\, \alpha^A_2\, D_Af+\frac{1}{2}\partial_u\big(\alpha^A_2\, D_A\alpha^B_2\big)\, D_Bf +\alpha^A_2\, \partial_u\alpha^B_2\, D_AD_Bf\label{lambda2var}
\end{aligned}
\end{equation}
with $\mathcal{L}_{Y}\lambda_2=Y^A\partial_A\lambda_2$.
By demanding the Einstein equations as in \cite{CJ2}, we can write 
\beq\label{intlambda2}
\lambda_2=\lambda_2^0(z,\zbr)+u\ \lambda_2^1(z,\zbr)+\Lambda_2(u,z,\zbr)
\eeq
where the form of $\Lambda_2(u,z,\zbr)$ will not be important in what follows. 
This leads to 
\begin{equation}
    \begin{aligned}
    \delta\lambda_2^0 &= \big[\psi+\mathcal{L}_{Y}\big]\lambda_2^0+ T\, \lambda_2^1-\tilde{\tau}-\frac{1}{4}\, a_3^A\, D_A\psi +\big(D_{\zbr}c^{z\zbr}- D_{z}c^{zz}+\gamma^{z\zbr}D_zD_{\zbr} a^z_2- \gamma^{z\zbr}D^2_{\zbr}a^{\zbr}_2\big)\, D_zT \\&+\big(D_{z}c^{z\zbr}- D_{\zbr}c^{\zbr\zbr}+\gamma^{z\zbr}D_{\zbr}D_z a^{\zbr}_2-\gamma^{z\zbr}D^2_z a^z_2\big)\, D_{\zbr}T+a^A_2D_A\xi^v_{(0)}+a^A_2D_AT +\frac{1}{4}a^A_2D_A\big(a^B_2D_B\psi\big)
    \end{aligned}
\end{equation}
The inhomogeneous part of this is the independent subleading hypertranslation, which takes the form
\begin{equation}
    \begin{aligned}
    \tilde{\tau} &=\tau-\frac{1}{4}\, a_3^A\, D_A\psi +\big(D_{\zbr}c^{z\zbr}- D_{z}c^{zz}+\gamma^{z\zbr}D_zD_{\zbr}a^z_2- \gamma^{z\zbr}D^2_{\zbr}a^{\zbr}_2\big)\, D_zT \\&+\big(D_{z}c^{z\zbr}- D_{\zbr}c^{\zbr\zbr}+\gamma^{z\zbr}D_{\zbr}D_z a^{\zbr}_2-\gamma^{z\zbr}D^2_z a^z_2\big)\, D_{\zbr}T +a^A_2D_A\xi^v_{(0)}+a^A_2D_AT +\frac{1}{4}a^A_2D_A\big(a^B_2D_B\psi\big)
    \end{aligned}
\end{equation}
This results in the modified algebra we presented earlier. 


\section{Covariant Surface Charges}\label{covaurface}

We will compute the covariant surface charges of \cite{Iyer} as in \cite{CJ2}, see also \cite{Barnich2}. For the set up in the present paper putting all the ingredients together leads to a potentially divergent term
\beq\label{doublenullcharge1}
\slashed{\delta}\mathcal{Q}_{\xi}[h;g] = \frac{1}{16\pi G}\lim_{v\to\infty}\int d^2\Omega\, \bigg[\frac{1}{8}\Big(\psi\, D_A\delta\alpha^A_2-2\, Y_A\, \delta\alpha^A_2-\psi\, \gamma^{AB} \delta\mathcal{C}_{AB}-\delta\alpha^A_2  D_A\psi-Y_A\, \partial_u\delta\alpha^A_3\Big)\, v+O\big(v^{0}\big)\bigg]\eeq
This should be compared to eqn. (54) of \cite{CJ2}. 
Substituting
\begin{subequations}
    \begin{align}
        \partial_u\delta\alpha^z_3 &= 2\, D_{\zbr}\delta\mathcal{C}^{z\zbr}-2\, D_z\delta\mathcal{C}^{zz}+2\, \gamma^{z\zbr}D_zD_{\zbr}\delta\alpha^z_2-2\, \gamma^{z\zbr}D_{\zbr}D_{\zbr}\delta\alpha^{\zbr}_2 \\
        \partial_u\delta\alpha^{\zbr}_3 &= 2\, D_z\delta\mathcal{C}^{z\zbr}-2\, D_{\zbr}\delta\mathcal{C}^{\zbr\zbr}+2\, \gamma^{z\zbr}D_zD_{\zbr}\delta\alpha^{\zbr}_2-2\, \gamma^{z\zbr}D_zD_z\delta\alpha^z_2
    \end{align}
\end{subequations}
we obtain
\beq\begin{aligned}\label{doublenullcharge1}
\slashed{\delta}\mathcal{Q}_{\xi}[h;g] = \frac{1}{16\pi G}\lim_{v\to\infty}\int d^2\Omega\, \bigg[
&\frac{1}{4}\gamma^{z\zbr}\, \Big(Y^z\big(D_{\zbr}\delta\mathcal{C}_{zz}-D_z\delta\mathcal{C}_{z\zbr}\big)+Y^{\zbr}\big(D_{z}\delta\mathcal{C}_{\zbr\zbr}-D_{\zbr}\delta\mathcal{C}_{z\zbr}\big)-\psi\, \delta\mathcal{C}_{z\zbr}\Big) \\&+\frac{1}{4}\Big(Y^z\, D_zD_z\delta\alpha^z_2-\delta\alpha^z_2\, D_z\psi+\frac{1}{2}D_z\big(\psi\, \delta\alpha^z_2\big)-Y^{\zbr}\, D_zD_{\zbr}\delta\alpha^z_2-\gamma_{z\zbr}Y^{\zbr}\, \delta\alpha^z_2 \\&+Y^{\zbr}\, D_{\zbr}D_{\zbr}\delta\alpha^{\zbr}_2-\delta\alpha^{\zbr}_2\, D_{\zbr}\psi+\frac{1}{2}D_{\zbr}\big(\psi\, \delta\alpha^{\zbr}_2\big)-Y^{z}\, D_{\zbr}D_{z}\delta\alpha^{\zbr}_2-\gamma_{z\zbr}Y^{z}\, \delta\alpha^{\zbr}_2\Big)\, v+O\big(v^{0}\big)\bigg]\end{aligned}\eeq
We will establish finiteness of the charges by showing that the $O\big(v\big)$ term vanishes. The terms on the first line in the parenthesis at $O\big(v\big)$ in the above expression can be rewritten as
\beq\begin{aligned}
    Y^z\big(D_{\zbr}\delta\mathcal{C}_{zz}-D_z\delta\mathcal{C}_{z\zbr}\big) &+Y^{\zbr}\big(D_{z}\delta\mathcal{C}_{\zbr\zbr}-D_{\zbr}\delta\mathcal{C}_{z\zbr}\big)-\psi\, \delta\mathcal{C}_{z\zbr} \\ &= Y^z\, D_{\zbr}\delta\mathcal{C}_{zz}+Y^{\zbr}\, D_{z}\delta\mathcal{C}_{\zbr\zbr}-Y^{z}D_z\delta\mathcal{C}_{z\zbr}-Y^{\zbr}D_{\zbr}\delta\mathcal{C}_{z\zbr}-\big(D_zY^z+D_{\zbr}Y^{\zbr}\big)\delta\mathcal{C}_{z\zbr}\\
    &=D_{\zbr}\big(Y^z\, \delta\mathcal{C}_{zz}\big)+D_{z}\big(Y^{\zbr}\, \delta\mathcal{C}_{\zbr\zbr}\big)-D_z\big(Y^z\, \delta\mathcal{C}_{z\zbr}\big)-D_{\zbr}\big(Y^{\zbr}\, \delta\mathcal{C}_{z\zbr}\big)\\ &=D_z\big(Y^{\zbr}\, \delta\mathcal{C}_{\zbr\zbr}-Y^z\, \delta\mathcal{C}_{z\zbr}\big)+D_{\zbr}\big(Y^z\, \delta\mathcal{C}_{zz}-Y^{\zbr}\, \delta\mathcal{C}_{z\zbr}\big)
\end{aligned}\eeq
Similarly, the terms on the second line in the parenthesis at $O\big(v\big)$ can be rewritten as
\beq\begin{aligned}
Y^z\, D_zD_z&\delta\alpha^z_2-\delta\alpha^z_2\, D_z\psi+\frac{1}{2}D_z\big(\psi\, \delta\alpha^z_2\big)-Y^{\zbr}\, D_zD_{\zbr}\delta\alpha^z_2-\gamma_{z\zbr}Y^{\zbr}\, \delta\alpha^z_2 \\ &= Y^z\, D_zD_z\delta\alpha^z_2-\delta\alpha^z_2\, D_zD_zY^z-\delta\alpha^z_2\, D_zD_{\zbr}Y^{\zbr}+\frac{1}{2}D_z\big(\psi\, \delta\alpha^z_2\big)-Y^{\zbr}\, D_zD_{\zbr}\delta\alpha^z_2-\gamma_{z\zbr}Y^{\zbr}\, \delta\alpha^z_2 \\ &= \big(Y^z\, D_zD_z\delta\alpha^z_2+D_z\delta\alpha^z_2\, D_zY^z\big)-D_z\big(\delta\alpha^z_2\, D_zY^z\big)-\delta\alpha^z_2\, D_zD_{\zbr}Y^{\zbr}+\frac{1}{2}D_z\big(\psi\, \delta\alpha^z_2\big)-Y^{\zbr}\, D_zD_{\zbr}\delta\alpha^z_2-\gamma_{z\zbr}Y^{\zbr}\, \delta\alpha^z_2 \\ &= D_z(Y^z D_z\delta\alpha^z_2)-D_z\big(\delta\alpha^z_2 D_zY^z\big)-\big(\delta\alpha^z_2\, D_{\zbr}D_zY^{\zbr}-\gamma_{z\zbr}\, \delta\alpha^z_2\, Y^{\zbr}\big)+\frac{1}{2}D_z\big(\psi\, \delta\alpha^z_2\big)+\Big(D_zY^{\zbr}\, D_{\zbr}\delta\alpha^z_2 \\& -D_z\big(Y^{\zbr} D_{\zbr}\delta\alpha^z_2\big)\Big)-\gamma_{z\zbr}Y^{\zbr}\, \delta\alpha^z_2 \\ &= D_z(Y^z D_z\delta\alpha^z_2)-D_z\big(\delta\alpha^z_2 D_zY^z\big)+\frac{1}{2}D_z\big(\psi\, \delta\alpha^z_2\big)-D_z\big(Y^{\zbr} D_{\zbr}\delta\alpha^z_2\big)
\end{aligned}\eeq
Note that in writing down the third equality in the above expression, we have commuted the covariant derivatives acting on $Y^{\zbr}$ using the definition of the Riemann tensor. That is, we have evaluated $[D_A,D_B]Y^C= R^C_{\, \, \, DAB}Y^D$ to obtain $D_zD_{\zbr}Y^{\zbr}-D_{\zbr}D_zY^{\zbr}=-\gamma_{z\zbr}Y^{\zbr}$. To simplify and obtain the final expression, we have made use of the fact that $Y^z$ and $Y^{\zbr}$ are holomorphic functions of $z$ and $\zbr$ respectively. A similar procedure can be implemented for $\delta\alpha^{\zbr}_2$ to rewrite the terms on the third line in the parenthesis at $O\big(v\big)$ as follows
\beq\begin{aligned}
Y^{\zbr}\, D_{\zbr}D_{\zbr}\delta\alpha^{\zbr}_2-\delta\alpha^{\zbr}_2\, D_{\zbr}\psi&+\frac{1}{2}D_{\zbr}\big(\psi\, \delta\alpha^{\zbr}_2\big)-Y^{z}\, D_{\zbr}D_{z}\delta\alpha^{\zbr}_2-\gamma_{z\zbr}Y^{z}\, \delta\alpha^{\zbr}_2\\ &= D_{\zbr}(Y^{\zbr} D_{\zbr}\delta\alpha^{\zbr}_2)-D_{\zbr}\big(\delta\alpha^{\zbr}_2 D_{\zbr}Y^{\zbr}\big)+\frac{1}{2}D_{\zbr}\big(\psi\, \delta\alpha^{\zbr}_2\big)-D_{\zbr}\big(Y^{z} D_{z}\delta\alpha^{\zbr}_2\big)
\end{aligned}\eeq
After integration over the 2-sphere, the ``total" derivative terms disappear. The vanishing of $O\big(v\big)$ terms guarantees that the surface charges  remain finite in the limit $v\rightarrow\infty$. This is one of our key results in this paper.

Due to the vanishing of the $O\big(v\big)$ terms, only the $O\big(v^0\big)$ terms remain in the $v \rightarrow \infty$ limit. These constitute our charge expression and they can be evaluated to be
\beq\begin{aligned}
\slashed{\delta}\mathcal{Q}_{\xi}[h;g] &=\frac{1}{16\pi G}\int d^2\Omega\, \Big[u\, Y_A\, \delta\alpha^A_2-\frac{3}{8}Y_A\, \delta\alpha^A_3-\frac{1}{4}Y^A \alpha^B_2\, \delta\mathcal{C}_{AB}-\frac{f}{2}\, \gamma^{z\zbr}\, \delta\mathcal{C}_{z\zbr}-\frac{1}{4}\gamma_{AB}\, \xi^A_{(1)}\, \delta\alpha^B_2-\frac{\psi}{8}\gamma_{AB}\, \alpha^A_2\, \delta\alpha^B_2 \\&-\frac{\psi}{4}\delta\lambda_2+u\, \frac{\psi}{2}\, \gamma^{z\zbr}\delta\mathcal{C}_{z\zbr}-\frac{3}{8}\psi\, \gamma^{z\zbr}\mathcal{D}_{z\zbr}-\frac{1}{4}Y^A\, \delta\alpha^B_2\, \mathcal{C}_{AB}+\frac{3}{16}\psi\, \delta\mathcal{C}^{AB}\, \mathcal{C}_{AB}+\psi\, \gamma^{AC}\, \delta\mathcal{C}_{AB}\, D_C\alpha^B_2 +\frac{f}{4}D_A\delta\alpha^A_2 \\&-\frac{1}{4}\xi^v_{(0)}\, D_A\delta\alpha^A_2-\frac{u}{4}\psi\, D_A\delta\alpha^A_2+\frac{1}{4}\delta\alpha^A_2\, D_A\xi^v_{(0)}+\frac{u}{4}\, \delta\alpha^A_2\, D_A\psi-\frac{1}{8}\delta\alpha^A_3\, D_A\psi+\frac{\psi}{8}\gamma^{z\zbr}\mathcal{C}_{z\zbr}\, D_A\delta\alpha^A_2 \\&+\frac{1}{4}\gamma^{z\zbr}\, \psi\, \alpha^A_2\, D_A\delta\mathcal{C}_{z\zbr}-\frac{1}{4}\gamma^{z\zbr}\delta\mathcal{C}_{z\zbr}\, \alpha^A_2\, D_A\psi-\frac{1}{8}\gamma^{z\zbr}\mathcal{C}_{z\zbr}\, \delta\alpha^A_2\, D_A\psi+\frac{1}{8}\gamma^{z\zbr}\, \psi\, \delta\alpha^A_2\, D_A\mathcal{C}_{z\zbr}-\frac{1}{4}\delta\alpha^A_2\, D_Af \\&+\frac{\psi}{8}D_A\delta\alpha^A_3+u\, \frac{\psi}{4}\gamma^{z\zbr}\delta\mathcal{C}_{z\zbr}-\frac{\psi}{8}\gamma^{z\zbr}\delta\mathcal{D}_{z\zbr}-\frac{1}{8}Y^A\, \delta\mathcal{C}_{AB}\, \partial_u\alpha^B_3+\frac{u}{2}\, Y_A\, \partial_u\delta\alpha^A_3+\frac{f}{2}\gamma^{z\zbr}\, \partial_u\delta\mathcal{D}_{z\zbr}-\frac{f}{8}\mathcal{N}^{AB}\, \delta\mathcal{C}_{AB} \\&-\frac{1}{8}\gamma_{AB}\, \xi^A_{(1)}\, \partial_u\delta\alpha^B_3-\frac{\psi}{16}\, \gamma_{AB}\, \alpha^A_2\, \partial_u\delta\alpha^B_3-\frac{1}{8}Y^A\, \mathcal{C}_{AB}\, \partial_u\delta\alpha^B_3-\frac{1}{8} Y_A\, \partial_u\delta\alpha^A_4-\frac{f}{4}\mathcal{C}^{AB}\, \delta\mathcal{N}_{AB}\Big]
\end{aligned}\eeq
Further on, rearranging the terms and simplifying the above expression gives
\beq\begin{aligned}
    \slashed{\delta}\mathcal{Q}_{\xi}[h;g] &=\frac{1}{16\pi G}\int d^2\Omega\, \Big[Y_A \Big(u\, \delta\alpha^A_2-\frac{3}{8}\delta\alpha^A_3+\frac{u}{2}\partial_u\delta\alpha^A_3-\frac{1}{8}\partial_u\delta\alpha^A_4-\frac{1}{4}\delta\big(\mathcal{C}^A_{\ \ B}\, \alpha^B_2\big)-\frac{1}{8}\delta\big(\mathcal{C}^A_{\ \ B}\, \partial_u\alpha^B_3\big)\Big)\\& +\psi\Big(-\frac{u}{2}D_A\delta\alpha^A_2+\frac{1}{4}D_A\delta\alpha^A_3+\frac{3}{4}u\, \gamma^{z\zbr}\, \delta\mathcal{C}_{z\zbr}-\frac{1}{2}\, \gamma^{z\zbr}\, \delta\mathcal{D}_{z\zbr}-\frac{1}{4}\delta\lambda_2-\frac{1}{8}\gamma_{AB}\, \alpha^A_2\, \delta\alpha^B_2 \\&+\frac{3}{16}\mathcal{C}^{AB}\, \delta\mathcal{C}_{AB}+D^A\alpha^B_2\, \delta\mathcal{C}_{AB}+\frac{1}{4}\gamma^{z\zbr}\, \alpha^A_2\, D_A\delta\mathcal{C}_{z\zbr}+\frac{1}{4}\delta\big(D_A(\gamma^{z\zbr}\, \mathcal{C}_{z\zbr}\, \alpha^A_2)\big)-\frac{1}{16}\gamma_{AB}\, \alpha^A_2\, \partial_u\delta\alpha^B_3\Big) \\&+f \Big(\frac{1}{2}D_A\delta\alpha^A_2-\frac{1}{2}\gamma^{z\zbr}\, \delta\mathcal{C}_{z\zbr}+\frac{1}{2}\gamma^{z\zbr}\, \partial_u\delta\mathcal{D}_{z\zbr}-\frac{1}{8}\mathcal{N}^{AB}\, \delta\mathcal{C}_{AB}-\frac{1}{4}\, \mathcal{C}^{AB}\, \delta\mathcal{N}_{AB}\Big)\\&+\xi^A_{(1)}\Big(-\frac{1}{4}\gamma_{AB}\, \delta\alpha^B_2-\frac{1}{8}\gamma_{AB}\, \partial_u\delta\alpha^B_3\Big)+\xi^v_{(0)}\Big(-\frac{1}{2}D_A\delta\alpha^A_2\Big)\Big]
\end{aligned}\eeq
Next we can substitute in the shifts that we obtained earlier, namely
\begin{subequations}
    \begin{align}
        \xi^A_{(1)} &= X^A-2\, D^Af\\
        \xi^v_{(0)} &= \phi+f+\Delta_{\gamma}f+\frac{u}{2}\psi-\frac{1}{2}D_AX^A-\frac{1}{4}\alpha^A_2\, D_A\psi
    \end{align}
\end{subequations}
to write down the final form of the charge expression as follows:
\beq\begin{aligned}
    \slashed{\delta}\mathcal{Q}_{\xi}[h;g] &=\frac{1}{16\pi G}\int d^2\Omega\, \Big[Y_A \Big(u\, \delta\alpha^A_2-\frac{3}{8}\delta\alpha^A_3+\frac{u}{2}\partial_u\delta\alpha^A_3-\frac{1}{8}\partial_u\delta\alpha^A_4-\frac{1}{4}\delta\big(\mathcal{C}^A_{\ \ B}\, \alpha^B_2\big)-\frac{1}{8}\delta\big(\mathcal{C}^A_{\ \ B}\, \partial_u\alpha^B_3\big)\Big)\\& +\psi\Big(-\frac{3}{4}u\, D_A\delta\alpha^A_2+\frac{1}{4}D_A\delta\alpha^A_3+\frac{3}{4}u\, \gamma^{z\zbr}\, \delta\mathcal{C}_{z\zbr}-\frac{1}{2}\, \gamma^{z\zbr}\, \delta\mathcal{D}_{z\zbr}-\frac{1}{4}\delta\lambda_2-\frac{1}{8}\gamma_{AB}\, \alpha^A_2\, \delta\alpha^B_2 +\frac{3}{16}\mathcal{C}^{AB}\, \delta\mathcal{C}_{AB}\\&+D^A\alpha^B_2\, \delta\mathcal{C}_{AB}+\frac{1}{4}\gamma^{z\zbr}\, \alpha^A_2\, D_A\delta\mathcal{C}_{z\zbr}+\frac{1}{4}\delta\big(D_A(\gamma^{z\zbr}\, \mathcal{C}_{z\zbr}\, \alpha^A_2)\big)-\frac{1}{16}\gamma_{AB}\, \alpha^A_2\, \partial_u\delta\alpha^B_3-\frac{1}{8}D_A\big(\alpha^A_2\, D_B\delta\alpha^B_2\big)\Big) \\&+f \Big(-\frac{1}{2}D_A\delta\alpha^A_2-\frac{1}{2}\gamma^{z\zbr}\, \delta\mathcal{C}_{z\zbr}+\frac{1}{2}\gamma^{z\zbr}\, \partial_u\delta\mathcal{D}_{z\zbr}-\frac{1}{4}\partial_uD_A\delta\alpha^A_3-\frac{1}{8}\mathcal{N}^{AB}\, \delta\mathcal{C}_{AB}-\frac{1}{4}\, \mathcal{C}^{AB}\, \delta\mathcal{N}_{AB}\Big)\\&-\frac{1}{2}\Delta_{\gamma}f\, D_A\delta\alpha^A_2+X_A\Big(-\frac{1}{4}\delta\alpha^A_2-\frac{1}{8}\partial_u\delta\alpha^A_3-\frac{1}{4}D^AD_B\delta\alpha^B_2\Big)+\phi\Big(-\frac{1}{2}D_A\delta\alpha^A_2\Big)\Big]
\end{aligned}\eeq
This can be expanded further by substituting the Einstein constraints on the metric parameters
\begin{subequations}
    \begin{align}
        \partial_u\alpha^z_3 &= 2\, D_{\zbr}\mathcal{C}^{z\zbr}-2\, D_z\mathcal{C}^{zz}+2\, \gamma^{z\zbr}D_zD_{\zbr}\alpha^z_2-2\, \gamma^{z\zbr}D_{\zbr}D_{\zbr}\alpha^{\zbr}_2 \\
        \partial_u\alpha^{\zbr}_3 &= 2\, D_z\mathcal{C}^{z\zbr}-2\, D_{\zbr}\mathcal{C}^{\zbr\zbr}+2\, \gamma^{z\zbr}D_zD_{\zbr}\alpha^{\zbr}_2-2\, \gamma^{z\zbr}D_zD_z\alpha^z_2
    \end{align}
\end{subequations}
but we will not do so here. 

The key observation we take away from the final form of the charges is that both leading hypertranslations and leading hyperrotations show up in these charges. This should be contrasted to our previous paper \cite{CJ2} where only the metric parameters corresponding to these diffeomorphisms showed up, and not the diffeomorphisms themselves. We will investigate the physical significance of hypertranslations and their connections to new memory effects in follow up work.


\vspace{-0.2in}

\begin{thebibliography}{99}

\bibitem{Witten}
E.~Witten,
``Anti-de Sitter space and holography,''
Adv. Theor. Math. Phys. \textbf{2}, 253-291 (1998)
doi:10.4310/ATMP.1998.v2.n2.a2
[arXiv:hep-th/9802150 [hep-th]].

\bibitem{dS} See eg., V.~Balasubramanian, J.~de Boer and D.~Minic,
``Notes on de Sitter space and holography,''
Class. Quant. Grav. \textbf{19}, 5655-5700 (2002)
doi:10.1016/S0003-4916(02)00020-9
[arXiv:hep-th/0207245 [hep-th]], and the first few references therein.

\bibitem{Bondi}
H.~Bondi, M.~G.~J.~van der Burg and A.~W.~K.~Metzner,
``Gravitational waves in general relativity. 7. Waves from axisymmetric isolated systems,''
Proc. Roy. Soc. Lond. A \textbf{269}, 21-52 (1962)
doi:10.1098/rspa.1962.0161.
R.~K.~Sachs,
``Gravitational waves in general relativity. 8. Waves in asymptotically flat space-times,''
Proc. Roy. Soc. Lond. A \textbf{270}, 103-126 (1962)
doi:10.1098/rspa.1962.0206; R.~Sachs,
``Asymptotic symmetries in gravitational theory,''
Phys. Rev. \textbf{128}, 2851-2864 (1962)
doi:10.1103/PhysRev.128.2851

\bibitem{Budhaditya}
B.~Bhattacharjee and C.~Krishnan,
``A General Prescription for Semi-Classical Holography,''
[arXiv:1908.04786 [hep-th]].

\bibitem{ACD}
C.~Krishnan,
``Bulk Locality and Asymptotic Causal Diamonds,''
SciPost Phys. \textbf{7}, no.4, 057 (2019)
doi:10.21468/SciPostPhys.7.4.057
[arXiv:1902.06709 [hep-th]].

\bibitem{Jude}
C.~Krishnan, V.~Patil and J.~Pereira,
``Page Curve and the Information Paradox in Flat Space,''
[arXiv:2005.02993 [hep-th]].

\bibitem{CJ1}
C.~Krishnan and J.~Pereira,
``A New Gauge for Asymptotically Flat Spacetime,''
[arXiv:2112.11440 [hep-th]].

\bibitem{WaldText}
R.~M.~Wald,
``General Relativity,''
Chicago Univ. Pr., 1984,
doi:10.7208/chicago/9780226870373.001.0001

\bibitem{Israel}
P.~R.~Brady, S.~Droz, W.~Israel and S.~M.~Morsink,
``Covariant double null dynamics: (2+2) splitting of the Einstein equations,''
Class. Quant. Grav. \textbf{13}, 2211-2230 (1996)
doi:10.1088/0264-9381/13/8/015
[arXiv:gr-qc/9510040 [gr-qc]].

\bibitem{Dafermos}
M.~Dafermos, G.~Holzegel and I.~Rodnianski,
``The linear stability of the Schwarzschild solution to gravitational perturbations,''
Acta Math. \textbf{222}, 1-214 (2019)
doi:10.4310/ACTA.2019.v222.n1.a1
[arXiv:1601.06467 [gr-qc]].




\bibitem{Strominger}
A.~Strominger,
``On BMS Invariance of Gravitational Scattering,''
JHEP \textbf{07}, 152 (2014)
doi:10.1007/JHEP07(2014)152
[arXiv:1312.2229 [hep-th]].

\bibitem{CJ2}
C.~Krishnan and J.~Pereira,
``Hypertranslations and Hyperrotations,''
[arXiv:2205.01422 [hep-th]].


\bibitem{Barnich}
G.~Barnich and C.~Troessaert,
``Aspects of the BMS/CFT correspondence,''
JHEP \textbf{05}, 062 (2010)
doi:10.1007/JHEP05(2010)062
[arXiv:1001.1541 [hep-th]].

\bibitem{Iyer}
V.~Iyer and R.~M.~Wald,
``Some properties of Noether charge and a proposal for dynamical black hole entropy,''
Phys. Rev. D \textbf{50}, 846-864 (1994)
doi:10.1103/PhysRevD.50.846
[arXiv:gr-qc/9403028 [gr-qc]].

\bibitem{Brandt}
G.~Barnich and F.~Brandt,
``Covariant theory of asymptotic symmetries, conservation laws and central charges,''
Nucl. Phys. B \textbf{633}, 3-82 (2002)
doi:10.1016/S0550-3213(02)00251-1
[arXiv:hep-th/0111246 [hep-th]].

\bibitem{StromingerR}
A.~Strominger,
``Lectures on the Infrared Structure of Gravity and Gauge Theory,''
[arXiv:1703.05448 [hep-th]].

\bibitem{CJRiemann} 
C.~Krishnan and J.~Pereira,
``Asymptotically Riemann-flat Spacetimes,'' to appear.

\bibitem{CJ}  C.~Krishnan and J.~Pereira,``A New Gauge for Flat Space Holography,'' to appear.

\bibitem{Schwarz}
J.~H.~Schwarz,
``Diffeomorphism Symmetry in Two Dimensions and Celestial Holography,''
[arXiv:2208.13304 [hep-th]].


\bibitem{Cachazo}
F.~Cachazo and A.~Strominger,
``Evidence for a New Soft Graviton Theorem,''
[arXiv:1404.4091 [hep-th]].


\bibitem{Barnich2}
G.~Barnich and C.~Troessaert,
``BMS charge algebra,''
JHEP \textbf{12}, 105 (2011)
doi:10.1007/JHEP12(2011)105
[arXiv:1106.0213 [hep-th]].

\end{thebibliography}
\end{document}